\documentclass[a4paper,fleqn, authoryear]{cas-dc}
\usepackage[round, authoryear]{natbib}
\usepackage{soul}
\usepackage{amsmath}
\usepackage{graphicx}
\usepackage{caption}
\definecolor{RED}{RGB}{156,78,90}

\def\tsc#1{\csdef{#1}{\textsc{\lowercase{#1}}\xspace}}
\tsc{WGM}
\tsc{QE}
\tsc{EP}
\tsc{PMS}
\tsc{BEC}
\tsc{DE}
\begin{document}
%\listoffigures
\captionsetup[figure]{labelfont={bf},labelformat={default},name={Fig.},labelsep=period}

\begin{sloppypar}

\let\WriteBookmarks\relax
\def\floatpagepagefraction{1}
\def\textpagefraction{.001}

\shorttitle{Ion strength–driven cavitation nucleation}
\shortauthors{J. Cai et~al.}

\title [mode = title]{Ionic strength–driven cavitation nucleation: from energy deposition-based to tension-based cavitation} 

\author[1]{Junhao Cai}
\fnmark[1]
\credit{Methodology, Validation, Investigation, Writing original draft, Writing-review $\&$ editing, Formal analysis, Data curation, Visualization}

\author[1]{Yuhan Li}
\fnmark[1]
\credit{Methodology, Software, Validation, Investigation, Writing original draft, Visualization}

\author[1,2]{Yunqiao Liu}
\credit{Supervision, Writing-Review \& Editing, Project administration}

\author[1,2]{Benlong Wang}
\credit{Supervision, Resources, Funding acquisition, Project administration}

\author[1,2]{Mingbo Li}
\cormark[1]
\cortext[cor1]{Corresponding author}
\credit{Conceptualization, Supervision, Methodology, Writing-Original Draft, Visualization, Funding acquisition, Project administration}
\ead{mingboli@sjtu.edu.cn}
\address[1]{School of Ocean and Civil Engineering, Shanghai Jiao Tong University, Shanghai 200240, China}
\address[2]{Key Laboratory of Hydrodynamics (Ministry of Education), Shanghai Jiao Tong University, Shanghai 200240, China}

\fntext[fn1]{Junhao Cai and Yuhan Li contributed equally to this work.}

\begin{abstract}
In this work, we present a unified experimental and simulation investigation of cavitation in aqueous electrolyte solutions, combining nanosecond laser–induced optical breakdown and all-atom molecular dynamics (MD) simulations under tensile stress. Across both cavitation scenarios, we find that cavitation inception and intensity (bubble nucleation count, cavitation-zone length, vapor-volume fraction) are governed by ionic strength alone, with negligible dependence on the ion species. In laser experiments, increasing ionic strength lowers the breakdown threshold and amplifies bubble generation by supplying extra seed electrons for inverse Bremsstrahlung–driven avalanche ionization. We elucidate the mechanism of action of the ionic strength through the MD simulations, which essentially quantifies the net charge density in the bulk, and thus its combined influence on the generation of seed electrons and the perturbation of the hydration network. These findings identify ionic strength serving as a unifying parameter controlling cavitation in electrolyte solutions: whether driven by rapid energy deposition or by tensile stress imposed. 
\end{abstract}

\begin{keywords}
Cavitation nucleation 
\sep Optical breakdown    
\sep Cavitation bubble 
\sep Ionic strength 
\sep Hydration network  
\end{keywords}

\maketitle

\section{Introduction}

Cavitation, the rapid formation and violent collapse of vapor or gas cavities in a liquid, plays a central role in fields ranging from marine propulsion to sonochemical reactors and biomedical ultrasound~\cite{brennen2014cavitation, stride2019nucleation, koukouvinis2021cavitation, nagalingam2023laser}. In practice, most of these applications involve electrolyte solutions (e.g., seawater, industrial effluents, physiological fluids), yet the extensive cavitation research has focused on pure water. This gap leaves critical uncertainties in predicting cavitation inception, erosion rates, and bubble dynamics under realistic ionic conditions. Addressing the influence of dissolved salt ions is therefore both scientifically necessary and urgently required to develop reliable models and control strategies for cavitation in real systems.

The onset and evolution of cavitation are governed by both the liquid's physicochemical properties and the local pressure field within the cavitation zone~\cite{supponen2016scaling, zeng2018wall}. In an ideal, impurity‐free liquid, cavitation occurs via homogeneous nucleation—a process whose fundamental principles were laid out by Gibbs and subsequently developed into modern nucleation theory~\cite{zeldovich1943theory}. Classical Nucleation Theory (CNT) predicts that pure water can sustain tensile stresses exceeding –140 MPa at ambient temperature before spontaneous vapor‐void formation occurs~\cite{PhysRevE.71.051605, kalikmanov2012classical}. Microscopically, thermal fluctuations generate transient low‐density "voids" that usually heal, but occasionally reach the critical radius required for stable bubble growth when assisted by mechanical or acoustic perturbations~\cite{zheng1991liquids, azouzi2013coherent}. In practice, however, experimental measurements consistently find cavitation thresholds far below CNT predictions, even in rigorously purified water, because heterogeneous nucleation on particles, microbubbles, or container surfaces drastically lowers the tensile strength~\cite{caupin2006cavitation, Berthelot, inclusions, lixinfa, shockwave, acousticcavi, gao2021effects, arora2004cavitation, arora2007effect, rossello2021demand}.

While extensive work has probed the role of microscopic or nanoscopic impurities in heterogeneous nucleation~\cite{bremond2006controlled, zhang2014controlled, saini2024finite}, the molecular‐scale influence of uniformly dissolved ions has received far less attention. Unlike solid or gas inclusions, dissolved ions distribute homogeneously yet subtly modify bulk properties—such as surface tension, gas solubility, and the electrostatic double layer—potentially shifting cavitation thresholds without offering discrete nucleation sites. A handful of studies on ionic liquids have hinted at altered cavitation and nanobubble stability in the presence of ions~\cite{johnson1981generation, brotchie2011role}, but a systematic understanding remains lacking.

To address this gap, we investigate the role of electrolyte ions in cavitation‐bubble nucleation through two complementary approaches: energy‐deposition–based laser‐induced breakdown experiments and tension‐based all-atom molecular dynamics (MD) simulations. Laser-induced cavitation, being a non-contact method, permits precise control over energy deposition and the isolation of specific factors (e.g., nanobubble populations~\cite{li2024effect}) without interference from extrinsic impurities. The ensuing explosive expansion generates shock waves and bubble oscillations that reveal the kinetics of plasma‐mediated cavitation. In parallel, MD simulations under imposed tensile stress elucidate how ionic strength perturbs the hydrogen‐bond network and lowers the mechanical barrier for bubble nucleation, providing atomistic insight into ion‐mediated cavitation dynamics. 

\begin{figure}
\centering
\includegraphics[width = 0.48\textwidth]{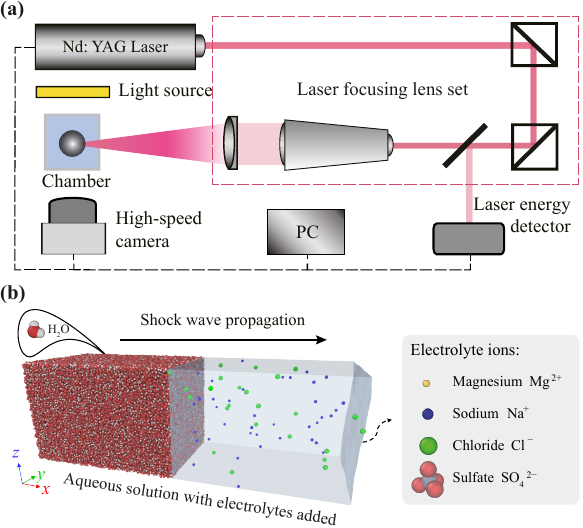}
\caption{(a) Experimental setup of laser-induced cavitation system.
(b) MD configuration of an aqueous solution system subjected to shock waves, where water molecules in the right half of the simulation box are folded to give a close-up of the soluble ions, exemplified by Na$^{+}$ and Cl$^-$. The right side shows four atomic configurations of ions embedded in the simulation box. }
\label{FIG1}
\end{figure} 

\begin{figure*}
\centering
\includegraphics[width = 1\textwidth]{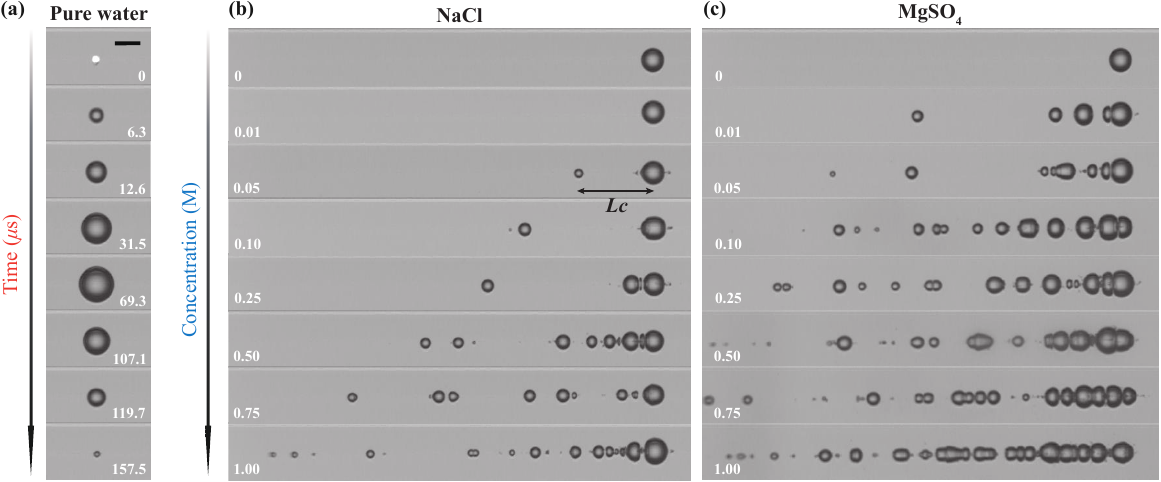}
\caption{Laser-induced breakdown and cavitation bubble in (a) ultrapure water, (b) NaCl and (c) $\rm{MgSO_{4}}$ solutions. The snapshots in NaCl and $\rm{MgSO_{4}}$ solutions were chosen at the moment when the cavitation bubbles near the focal point had almost reached their maximum size. The laser pulse energy density for all cases is 6.25 $\rm{J/cm^2}$. All snapshots share the same scale bar with a length of 2 $\rm{mm}$.} 
\label{FIG2} 
\end{figure*}

\section{Methods}

%% eletrolytes preparation
\textbf{Experimental.} 
All aqueous solutions, including control samples, were prepared using ultrapure water (resistivity: 18.2 M$\Omega$·cm, pH6.5) sourced from a Milli-Q purification system (Merck, Germany). Electrolyte solutions with molar concentrations ranging from 0.001 to 5 mol/L (M) for NaCl and 0.001 to 1 M for $\rm{MgSO_{4}}$ were prepared using analytical-grade salts (NaCl: $\geq$99.999$\%$, Thermo Scientific; $\rm{MgSO_{4}}$: $\geq$99.5$\%$, Sigma-Aldrich). Complete dissolution was achieved by magnetic stirring at 500 rpm (HS8Pro, JoanLab, China) for 10 min under temperature-controlled conditions (25$^\circ$C).

The experimental configuration for laser-induced cavitation and visualization is depicted in Figure~\ref{FIG1}(a). A nano-pulsed laser (Q-switched Nd: YAG, Dawa-100, 1064 nm wavelength, 9 ns pulse width, BeamTech, China) delivered high-energy pulses to induce optical breakdown within a quartz chamber (dimensions: 50 $\times$ 50 $\times$ 50 
mm$^3$) filled with either ultrapure water or electrolyte solutions. The chamber was securely mounted and sealed to minimize vibrations and prevent external contamination.
Laser pulse energies, ranging from 1.5 to $\rm{6.5\ J/cm^2}$, were measured using an energy meter (FieldMaxII, Coherent, USA) integrated into the beam path. The expanded laser beam was precisely focused at the geometric center of the chamber to mitigate boundary effects from the chamber walls and free liquid surface on cavitation dynamics. The focused laser pulse initiated localized optical breakdown, generating plasma that expanded to form a spherical vapor bubble. The progression from optical breakdown to cavitation inception was captured using a high-speed camera (S1310M, Revealer, China; 150,000 fps) . The recorded field of view ($\rm{42\times3.2\ mm^2}$) was imaged at a resolution of $1152\times88$ pixels per frame. Diffuse back-illumination was employed to enhance image contrast. Synchronization between the laser and camera was achieved using a pulse delay generator, ensuring precise temporal alignment.

\textbf{MD simulation details.} 
To elucidate the role of electrolytes in cavitation inception at a microscopic level, we developed an all-atom molecular dynamics model, as illustrated in Figure~\ref{FIG1}(b).
The computational domain comprises a rectangular aqueous solution chunk measuring $50\times15\times15$\ nm$^3$, containing over one million atoms.
This system includes two types of electrolytes, representing four ion species (Mg$^{2+}$, Na$^{+}$, Cl$^-$, and SO$_4^{2-}$), which matched those used in the experiment. Periodic boundary conditions were applied in all three spatial dimensions. The water molecules were modeled using the TIP4P/2005 potential, and ion parameters were adopted from force fields optimized for compatibility with this water model. Key parameters and additional details are provided in Supplemental Material (section 1).
The system underwent equilibrium relaxation under an isothermal-isobaric (NPT) ensemble at 298 K and 1 bar for 200 ps, employing a timestep of 1 fs to achieve a stable initial state.
Subsequently, the ensemble was converted into a microcanonical (NVE) ensemble to simulate the cavitation 
under a transient pressure fluctuation triggered by a shock wave propagation. 
Herein, the microcanonical ensemble has been widely used in the transient shock propagation scenarios to allow pressure and temperature fluctuate~\cite{zhan2021molecular, Rawat}.
%To initiate phase transition

This shock wave, in the present study, was generated by applying a constant-velocity displacement ($\sim$1.8 km/s) along the $x$-direction to the leftmost free surface of the liquid slab, mimicking a mobile "piston". This displacement continued until a total shift of $\sim$2.0 nm was achieved, after which the piston movement ceased, allowing the system to evolve freely for an additional 100 ps. Trajectories were recorded every 1.0 ps for analysis. The initial shock intensity was defined as $I_s=\rho_0 U_pU_s$, where $\rho_0$ is the initial liquid density, $U_p$ is the piston velocity, and $U_s$ is the resulting shock wave velocity. The relationship between $U_p$ and $U_s$ is detailed in FIG. S1 (see Supplemental Material), leading to an estimated shock intensity of approximately 8.15 GPa for this study.   

\section{Laser-induced cavitation: energy deposition-based}

Figure~\ref{FIG2} shows experimental snapshots of laser‑induced optical breakdown and subsequent cavitation activities in ultrapure water and in NaCl/$\rm MgSO_4$ electrolyte solutions. Under a laser energy density of $E = \rm {6.25\ J/cm^2}$—well above the cavitation threshold in ultrapure water—deterministic formation of a single, spherical bubble was achieved: plasma formation is followed by bubble expansion to a maximum radius of approximately $\sim$$700\ \mu$m. When the same laser parameters were applied to NaCl solutions at concentrations $C\geq0.05$ M, interestingly, multiple bubbles formed and propagated backward along the laser-focusing axis; both the number of cavitation bubbles and their travel distance increased with salt concentration. The absence of bubble formation on the right side of the focal point can be attributed to the plasma shielding effect~\cite{tian2016stabilization, fu2024secondary, fu2018experimental}. This results in strong energy absorption by the plasma, significantly reducing laser transmission through the region. In $\rm{MgSO_4}$ solutions, laser-induced breakdown can occur at lower electrolyte concentrations (0.01 M). Although all parameters remain the same, the cavitation activity is more intense with more cavitation bubbles formed along the laser-focusing axis. They are closely arranged on the optical axis and even merge with each other before they grow to their maximum size. The dynamic evolution of cavitation bubbles in solutions with different NaCl/$\rm{MgSO_4}$ concentrations can be seen in FIG. S2 (Supplemental Material). 

\begin{figure*}
\centering
\includegraphics[width = 0.98\textwidth]{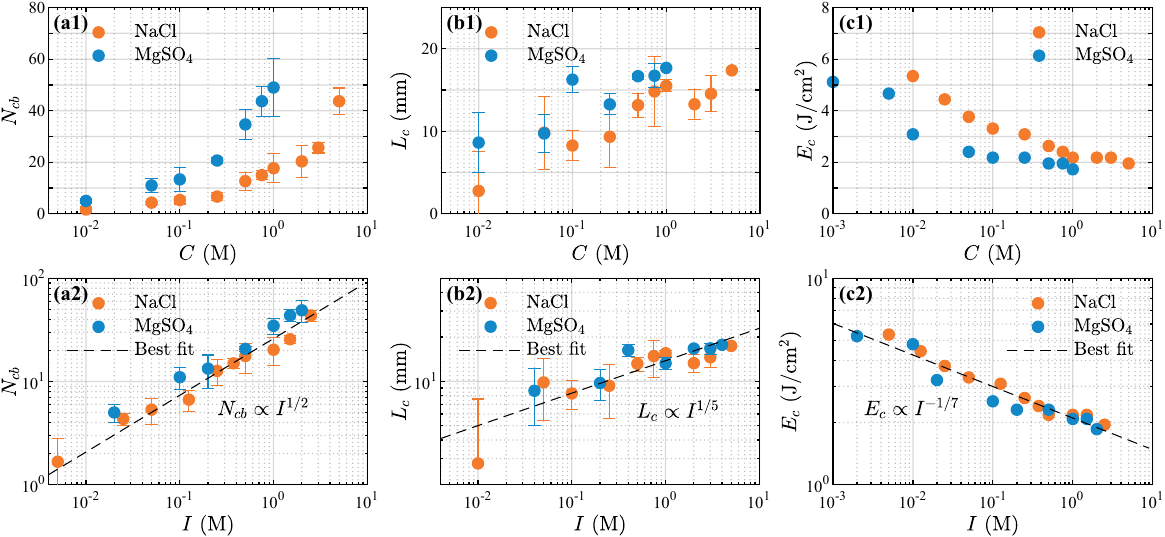}
\caption{(a1-a2) Number of nucleated cavitation bubbles $N_{cb}$ as a function of electrolyte concentration $C$ and corresponding ionic strength $I$. (b1-b2) Length of the cavitation zone along the laser focus axis $L_c$ as a function of electrolyte concentration $C$ and corresponding ionic strength $I$. (c1-c2) Laser breakdown threshold $E_c$ as a function of electrolyte concentration $C$ and corresponding ionic strength $I$. The dashed lines represent the power-law fit for the data from experiments. Error bars represent the standard error on the mean.}
\label{FIG3}
\end{figure*}

Quantitative analysis of laser‐induced, plasma‐mediated cavitation was carried out in NaCl (0$\sim$5 M) and $\rm{MgSO_4}$ (0$\sim$1 M) solutions under identical irradiation conditions. For each experiment, we recorded the total number of cavitation bubbles nucleated ($N_{cb}$) and the axial length of the cavitation zone ($L_c$)—metrics plotted in Figure~\ref{FIG3}(a1) and (b1). As electrolyte concentration increases, both $N_{cb}$ and $L_c$ rise monotonically, but $\rm{MgSO_4}$ consistently produces a higher nucleation density and a longer cavitation column than NaCl at equivalent molarities. This disparity suggests an ion-specific effect, likely rooted in the differing charge states of monovalent versus divalent ions. 
Besides, the first oscillation period ($T_{c}$) and maximum diameter ($D_{max}$) of the primary bubble at the focal point remain largely invariant across electrolyte concentrations, though $D_{max}$ exhibits a minor reduction at high concentrations (see FIG. S3). In ultrapure water, these parameters adhere to the Rayleigh collapse time~\cite{vogel1999energy}, $T_{c} = 0.915 D_{max} (\rho/(p_{s}-p_{v}))$, where $\rho$ represents the water density, $p_{s}$ the static pressure, and $p_{v}$ the vapor pressure. Multiple optical breakdown-induced bubble coalescence events, however, significantly prolong the bubble lifetime by up to 48$\%$ for comparable values, highlighting a distinct departure from classical cavitation behavior under ionic conditions (see FIG. S4).

To isolate charge‐related contributions, we reparameterized our data in terms of ionic strength,
\begin{equation}
\label{ionic}
I = \frac{1}{2} \sum c_i z_i^2 ,
\end{equation}
where $c_i$ is the molar concentration and $z_i$ the valence of ion species $i$. Remarkably, when $N_{cb}$ and $L_c$ are plotted against $I$ (Figure~\ref{FIG3}(a2) and (b2)), the data for both electrolytes collapse onto a single master curve, demonstrating that cavitation activity scales with ionic strength alone. In particular, the nucleation count follows a sublinear scaling law, $N_{cb} \propto I^{1/2}$, indicating that doubling the net charge density yields only a $\sim40\%$ increase in bubble events. 

We further quantified the optical breakdown threshold $E_c$, defined as the minimum energy density required to generate a dense thermal plasma and initiate explosive vaporization within the focal region. The optical breakdown comprehensively reflects the liquid's ability to resist damage under a strong laser field and is a comprehensive manifestation of its optical, thermal, electrical properties and chemical composition. Figure~\ref{FIG3}(c1) shows that $E_c$ decreases sharply with increasing electrolyte concentration, with $\rm{MgSO_4}$ solutions exhibiting consistently lower thresholds than NaCl. When plotted against $I$, $E_c$ universally follows an inverse power law, $E_c \propto I^{-1/7}$, as shown in Figure~\ref{FIG3}(c2). These results establish ionic strength as the key parameter governing laser‐induced cavitation dynamics—modulating both bubble nucleation statistics and the energy barrier for optical breakdown—irrespective of the specific ionic species present.

\begin{figure}
\centering
\includegraphics[width = 0.49\textwidth]{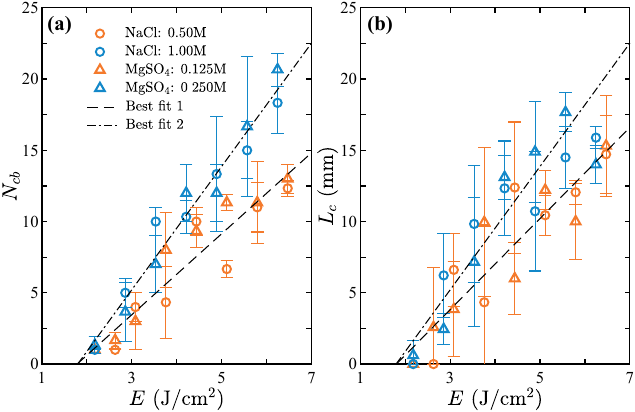}
\caption{(a) Number of nucleated cavitation bubbles $N_{cb}$ and (b) length of the cavitation zone along the laser focus axis $L_c$ as a function of laser pulse energy density $E$. Error bars represent the standard error on the mean. Both dashed and dotted lines represent the best linear fit.}
\label{FIG4}
\end{figure}

Physically, the addition of electrolyte ions mainly changes the thermal and electrical properties of the aqueous solutions~\cite{zaytsev1992properties}. While the addition of electrolyte ions does modify water’s thermophysical properties—raising the boiling point, altering thermal conductivity, and increasing the latent heat of vaporization (see FIG. S5, take NaCl solution as an example)—these changes primarily redistribute deposited laser energy among phase change, heat diffusion, and plasma formation and cannot fully account for the observed shifts in optical breakdown thresholds. For example, boiling‐point elevation increases the energy required for vapor nucleation; higher thermal conductivity promotes rapid heat removal, quenching local overheating and thus raising the fluence needed for breakdown; and greater latent heat of vaporization directly consumes more laser energy in creating vapor, further elevating the energy barrier to plasma initiation. However, these thermal‐based effects alone neither explain nor consistently match the measured breakdown and cavitation behavior.

Instead, the solution’s electrical properties dominate in the process. In the nanosecond regime, achieving optical breakdown in ultrapure water requires a sufficiently high laser energy density to generate an initial population of free electrons—initiated by multiphoton ionization and subsequently dominated by cascade (avalanche) ionization under inverse Bremsstrahlung absorption~\cite{goueguel2014effect}. When electrolytes are introduced, thermal excitation of dissolved ions—whose first ionization potentials are 5.14 eV for Na and 7.65 eV for Mg—provides additional seed electrons, thereby elevating the free electron density in the laser focal volume. These seed electrons then absorb laser energy via inverse Bremsstrahlung, gain kinetic energy, and collide with surrounding molecules to trigger an avalanche ionization~\cite{goueguel2015matrix}. The lower breakdown threshold observed in $\rm{MgSO_4}$ solutions arises from their higher free electron density, which strengthens electron–ion coupling and enhances avalanche ionization efficiency. Consequently, the intensity of cavitation activity correlates with the ionic strength, i.e., the free electron density, rather than simply the molar concentration of the electrolyte. 

\begin{figure*}
\centering
\includegraphics[width = 0.98\textwidth]{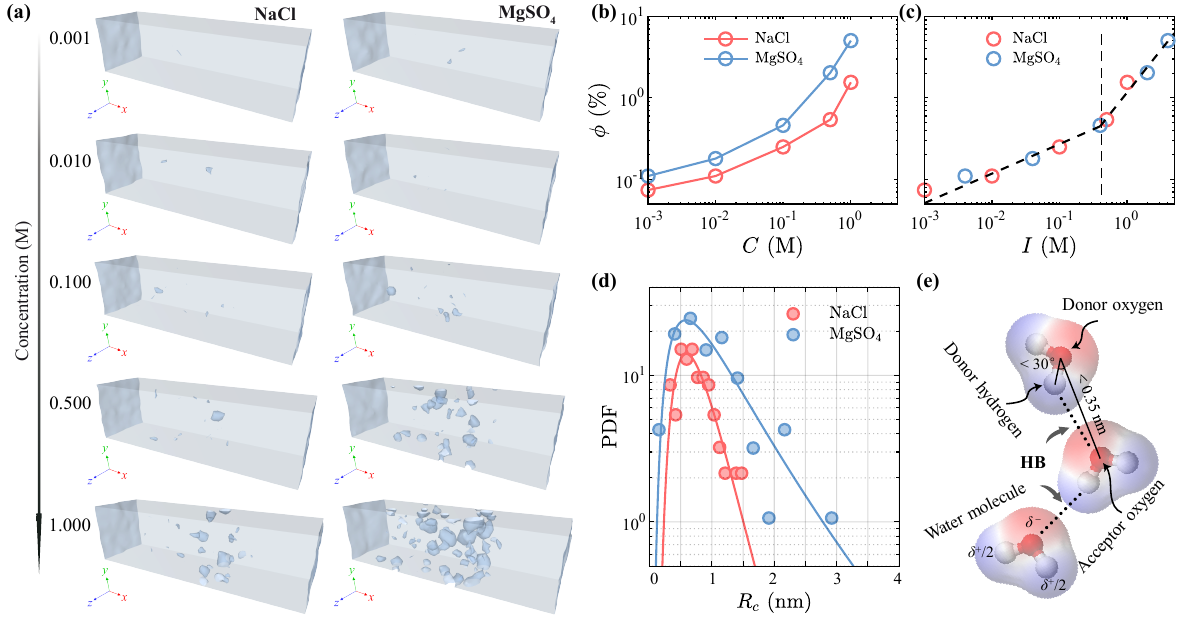}
\caption{(a) Snapshots of the MD trajectories capturing the vapor cavity nucleation for the NaCl and $\rm{MgSO_4}$ electrolyte solutions. 
Vapor volume fraction, $\phi$, at the inception stage in the two types of electrolyte solutions as a  functions of the (b) electrolyte concentration $C$ and (c) ionic strength $I$. (d) Size distribution of bulk vapour cavities in NaCl and $\rm{MgSO_4}$ solutions at the same ionic strength ($I$ = 1 M). The solid lines denote the fitting results with a log-normal distribution function. (e) Schematic diagrams of the hydrogen bond formation criterion among water molecules and of the dipole vector inside one specific water molecule. }
\label{FIG5}
\end{figure*}

To further validate the role of ionic strength $I$ in governing cavitation behavior and to assess how varying laser energy density $E$ influences bubble nucleation, we conducted a complementary series of experiments using NaCl and $\rm{MgSO_4}$ solutions prepared in a 4:1 molar ratio, thereby ensuring identical ionic strength in each solution pair. As shown in Figure~\ref{FIG4}, both $N_{cb}$ and $L_c$ increase linearly with the incident laser energy density $E$. Remarkably, under every energy condition tested, matched NaCl and $\rm{MgSO_4}$ solutions yield indistinguishable cavitation responses (see FIG. S6): their $N_{cb} (E)$ and $L_c(E)$ datasets collapse onto the same master curve. These findings confirm that, once $I$ is held constant, the primary determinant of bubble nucleation density and cavitation extent is the laser energy input, rather than the specific identity or valence of the dissolved ions. 

Moreover, the observed linear scaling suggests that each incremental increase in energy density contributes a fixed number of additional seed electrons. Experimental laser-induced breakdown spectroscopy (LIBS) studies have confirm that, above the breakdown threshold, the conduction‐band electron density $n_e$ scales linearly with laser fluence $F$ for nanosecond pulses in water up to moderate densities~\cite{varghese2015influence}. Under these conditions, the ratio $\Delta n_e/\Delta F$ remains constant, reflecting the fixed inverse Bremsstrahlung absorption and ionization probabilities per unit energy. In the avalanche‐dominated regime of nanosecond laser breakdown, each incremental increase in laser energy density yields a nearly constant addition of seed electrons because the inverse Bremsstrahlung absorption coefficient and electron–ion collision frequency remain effectively unchanged over the range of tested fluences~\cite{turnbull2023inverse}. Once the initial seed‐electron population, set by $I$, is established, absorbed energy is converted into kinetic energy of these electrons with a fixed probability of generating new electrons via impact ionization~\cite{park2019electron}. Consequently, the free‐electron density (and thus cavitation bubble nucleation) grows linearly with laser fluence until higher‐order effects such as plasma shielding become significant.

\section{MD results: Tension-based}

All-atom MD simulations were conducted to elucidate the role of dissolved ions in cavitation inception under a rapid pressure-relaxation protocol. In NaCl solutions, the simulated trajectories (see FIG. S7) reveal that cavitation events in the bulk emerge on the order of 70 ps after the instantaneous pressure drop. Visual inspection of simulation snapshots (see Figure~\ref{FIG5}(a)) indicates markedly enhanced cavity nucleation at higher electrolyte concentrations (especially above $\sim$0.1 M), with $\rm{MgSO_4}$ solutions yielding substantially more cavities than NaCl under comparable conditions. 

To quantify these observations, we compute the vapor-phase volume fraction $\phi$ at $t = 40\ \rm ps$ via an $\alpha$-shape algorithm, defining $\phi = 100\% \cdot\sum V_c/V_L$, where $V_c$ is the total volume of all identified cavities and $V_L$ is the volume of the aqueous solution. 
Here vapor volume fraction is preferred over bubble count for assessing cavitation nucleation, as it provides a dimensionless, intrinsic measure of nucleation extent. MD simulations enable comprehensive observation of cavitation throughout the simulation domain, allowing accurate computation of vapor volume fraction. In contrast, experimental setups, particularly those involving laser-induced cavitation, typically induce localized cavitation events along the laser's focal axis. This spatial confinement limits the applicability of volume fraction measurements in experimental contexts, as cavitation does not permeate the entire fluid volume. Therefore the cavitation bubble counting is applicable in experimental studies to account for the localized nature of cavitation events.
The results demonstrate that cavitation is driven by the tensile stresses imparted by the $x$-direction rarefaction wave, akin to a mechanical fracture of the liquid. The resulting $\phi$ values (see Figure~\ref{FIG5}(b)) reproduce the experimental trend (Figure~\ref{FIG3}(a)) with electrolyte concentration (noting that $\phi$ is a volume-based metric whereas the experiment measures cavitation-bubble count). Crucially, when $\phi$ is plotted against ionic strength $I$ (see Figure~\ref{FIG5}(c)), the data for NaCl and $\rm{MgSO_4}$ collapse onto a single master curve, indicating that ionic strength governs the cavitation response, which is surprisingly consistent with the conclusions obtained from laser-induced cavitation experiments. 

Further, $\phi$ vs. $I$ curve exhibits a marked "knee" at around 0.4 M, beyond which bubble yield rises steeply. There are two possible reasons for this transition. First, at moderate $I$ the hydration shells of neighbouring ions partially overlap but remain discrete. Once $I$ exceeds a threshold, these shells coalesce into a percolating, partially disordered hydrogen bond network, reducing the effective surface tension and energy barrier to nucleation, and thus promoting cavity growth and coalescence. Second, above a threshold $I$ the ions begin to form transient clusters, producing localised regions of enhanced structuring and conversely of weakened cohesion, where cavities preferentially nucleate. Besides, we also compared the scaling dependence of the number of cavitation bubbles $N_{cb}$ measured in experiments and MD simulations on the ionic strength $I$ (see FIG. S8). The results reveal a consistent order of magnitude for $N_{cb}$ across both methodologies, with similar quantitative relationships and comparable scaling-law exponents. We also calculated the size distribution of bulk vapor cavities in the two electrolyte solutions, and found that the size of the vapor cavities formed in the $\rm{MgSO_4}$ solution was larger, even twice that in the NaCl solution, as shown in Figure~\ref{FIG5}(d). In particular, the sizes of the vapor voids nucleated in the bulk are not randomly distributed but well follow a log-normal distribution. 

This behaviour can be understood in terms of ion-water interactions and disruption of the hydrogen bonding network in the bulk. Solvated ions perturb water’s hydrogen–bond (HB) network by forming tightly bound hydration shells that replace or reorient water–water hydrogen bonds. Figure~\ref{FIG5}(e) shows the sketch of the geometric hydrogen bond formation criterion that regulates Donor-Acceptor distance $r_{\rm{OO}}< 0.35$ nm and $\angle\rm{O-O-H} < 30^{\circ}$. Theoretical models, describing the interaction of ionic hydration shells with the hydrogen-bonding network of water, show that these hydration shells locally disrupt the H-bonding alignment of neighboring water molecules~\cite{macias2023hydrogen}. Actually, $\rm Mg^{2+}$ and $\rm SO_4^{2-}$ carry high charge density and bind water very strongly (kosmotropic behavior), whereas $\rm Na^{+}$ and $\rm Cl^{-}$ have lower charge density and weaker hydration (borderline chaotropes). For example, $\rm Mg^{2+}$ typically has a tightly bound first solvation shell of $\sim$6 water molecules, each strongly oriented, while $\rm SO_4^{2-}$ also coordinates a well‐structured shell. In contrast, $\rm Cl^-$ is larger and more polarizable, so its hydration shell is looser and its electrostatic field decays more rapidly. The observed differences in Figure~\ref{FIG5}(b) thus follow the classical Hofmeister ordering, resulting in the net effect: the water–water hydrogen-bond network is more perturbed near kosmotropes than near chaotropes. Indeed, analysis of water–water hydrogen bonding (see FIG. S9(a)) shows fewer hydrogen bonds in $\rm{MgSO_4}$ solutions than in NaCl at the same electrolyte concentration $C$.    

Although monovalent ions and divalent ions differ in their individual solvation dynamics, when the ionic strength $I$ is held constant, their macroscopic effects on cavitation coincide. This is because $I$ essentially quantifies the total charge density (the total charge per unit volume) in the bulk, and thus its combined influence on seed electron generation and hydration network perturbation. This is further supported by the fact that the number of hydrogen bonds possessed by each water molecule $N_h^*$ in different types of electrolyte solutions has the same scaling dependence on $I$ in the simulation results (see FIG. S9(b)). The total charge accumulation effect outweighs the differences in ionic species. Our previous works have shown that the behavior of nanobubbles or cavitation systems is mainly related to the accumulation of net charge in the solution rather than specific ion species~\cite{ma2022ion}. As ionic strength increases, more water molecules lie within these disrupted regions, weakening the overall network continuity. This loss of cohesion and tensile strength makes the liquid more susceptible to rupture under negative pressures, thereby facilitating nucleation of vapor cavities. 

\section{Conclusions}

Both our laser-induced cavitation experiments and MD simulations demonstrate that cavitation inception and intensity depend exclusively on ionic strength. Ionic strength governs: (1) the initial electron density and avalanche gain during optical breakdown and (2) the reduction in liquid cohesion via  hydrogen-bond disruption during nucleation. This dual control yields the observed single-parameter scaling: bubble number, cavitation zone length and vapor-volume fraction all increase uniformly with ionic strength, independent of ionic species. Elevated ionic strength lowers the breakdown threshold and boosts avalanche ionization efficiency, resulting in a greater number of cavitation "seed" electrons and plasma clusters during the optical breakdown stage. The greater cavitation in multivalent salt solutions arises from its kosmotropic ions' stronger distortion of water structure, effectively "seeding" heterogeneous nucleation and reducing the critical barrier. This detailed physical picture, linking ion-specific hydration, network disruption, and nucleation energetics, coherently extends the MD observations. Indeed, even other solution properties, such as surface tension, also exhibit linear dependencies on ionic strength irrespective of ion type~\cite{LI_ions}. Therefore, ionic strength, representing the solution’s overall charge environment, is the fundamental parameter unifying the photoionization efficiency and mechanical weakening that drive cavitation activity. The future work will focus on cavitation events in acidic or alkaline solutions, in particular inspired by our previous work: hydroxide ions $\rm OH^{-}$ can effectively promote the nucleation of nanobubbles and their stability.  

\printcredits

\section*{Declaration of Competing Interest}

The authors declare that they have no known competing financial interests or personal relationships that could have appeared to influence the work reported in this paper.

\section*{Acknowledgments}
This work is supported by National Natural Science Foundation of China under Grants Nos. 12202244 and 92252205, and the Oceanic Interdisciplinary Program of Shanghai Jiao Tong University (No: SL2023MS002). 

\appendix
\section*{Appendix A. Supplementary material}

Supplementary data associated with this article can be found in a separate file. 

%\bibliographystyle{elsarticle-num}

% \nocite{*}
% \bibliography{references}
%\bibliography{cas-refs}

\end{sloppypar}

\end{document}